\input mssymb.tex

 \def\CH{\hbox{{$\cal H$}}}

\def\CR{\hbox{{$\cal R$}}}

\def\R{{\Bbb R}}
\def\C{{\Bbb C}}
\def\Z{{\Bbb Z}}

\def\vecu{{\bf u}}

\def\sA{{\scriptstyle A}}
\def\sB{{\scriptstyle B}}
\def\sM{{\scriptstyle M}}
\def\grav{{\scriptstyle G}}
\def\<{\langle}
\def\>{\rangle}
\def\del{{\partial}}
\def\lform{\hbox{$\sqcup$}\llap{\hbox{$\sqcap$}}}

\def\h{{{1\over2}}}
\def\eps{{\epsilon}}

\def\rcross{{\triangleright\!\!\!<}}
\def\lcross{{>\!\!\!\triangleleft}}

\def\lcocross{{>\!\!\blacktriangleleft}}
\def\cobicross{{\triangleright\!\!\!\blacktriangleleft}}
\def\bicross{{\blacktriangleright\!\!\!\triangleleft}}
\def\dcross{{\bowtie}}

\def\nosum{{}}
\def\tens{\mathop{\otimes}}
\def\la{{\triangleright}}\def\ra{{\triangleleft}}
\def\isom{{\cong}}

\def\id{{\rm id}}
\def\Deltaop{{\Delta^{\rm op}}}

\def\proof{\goodbreak\noindent{\bf Proof\quad}}
\def\endproof{{\ $\lform$}\bigskip }

\def\s#1{{}_{\scriptscriptstyle(#1)}}
\def\o{{}_{\scriptscriptstyle(1)}}
\def\t{{}_{\scriptscriptstyle(2)}}
\def\th{{}_{\scriptscriptstyle(3)}}
\def\fo{{}_{\scriptscriptstyle(4)}}
\def\fiv{{}_{\scriptscriptstyle(5)}}
\def\six{{}_{\scriptscriptstyle(6)}}
\def\sev{{}_{\scriptscriptstyle(7)}}

\def\bo{{}^{\bar{\scriptscriptstyle(1)}}}
\def\bt{{}^{\bar{\scriptscriptstyle(2)}}}

\def\uo{{{}^{\scriptscriptstyle(1)}}}
\def\ut{{{}^{\scriptscriptstyle(2)}}}

\def\equad{\kern -1.7em}
\def\qqquad{\qquad\quad}
\def\nquad{{\!\!\!\!\!\!}}
\def\nqquad{\nquad\nquad}

\def\eqn#1#2{\begin{equation}#2\label{#1}\end{equation}}
\def\cmath#1{\[\begin{array}{c} #1 \end{array}\]}
\def\ceqn#1#2{\begin{equation}\label{#1}\begin{array}{c}#2
\end{array}\end{equation}}
\def\align#1{\begin{eqnarray*}#1\end{eqnarray*}}

\documentstyle[12pt,epsf]{article}
\newtheorem{lemma}{Lemma}[section]
\newtheorem{propos}[lemma]{Proposition}
\newtheorem{example}[lemma]{Example}
\newtheorem{theorem}[lemma]{Theorem}

\begin{document}\baselineskip 11pt

{\ }\hskip -.2in
{\em To appear:} Proc. Generalised Symmetries, Clausthal, Germany, July, 1993.
World Sci.
\vspace{.5in}

\begin{center} {\Large CROSS PRODUCT QUANTISATION, NONABELIAN\\{\ }\\
COHOMOLOGY AND TWISTING OF HOPF ALGEBRAS}\footnote{1991
Mathematics Subject Classification 16W30, 17B37, 55R10, 57T10, 58B30, 81R50,
81T70
$\qquad\qquad\qquad$
This paper is in final form and no version of it will be submitted for
publication elsewhere}
\\ \baselineskip 13pt{\ }
{\ }\\ S. Majid\footnote{Royal Society University Research Fellow and Fellow of
Pembroke College, Cambridge}\\ {\ }\\
Department of Applied Mathematics \& Theoretical Physics\\ University of
Cambridge, Cambridge CB3 9EW, U.K.
\end{center}

\vspace{10pt}
\begin{quote}\baselineskip 11pt
\noindent{\bf ABSTRACT} This is an introduction to work on the generalisation
to quantum groups
of Mackey's approach to quantisation on homogeneous spaces. We recall the
bicrossproduct models of the author, which generalise the quantum double. We
describe the general extension theory of Hopf algebras and the nonAbelian
cohomology spaces $\CH^2(H,A)$ which classify them. They form a new kind of
topological quantum number in physics which is visible only in the quantum
world. These same cross product quantisations can also be viewed as trivial
quantum principal bundles in quantum group gauge theory. We also relate this
nonAbelian cohomology $\CH^2(H,\C )$ to Drinfeld's theory of twisting.

\bigskip

\end{quote}
\baselineskip 15pt

\section{Introduction}

This paper is concerned with quantum algebras of observables that happen to be
Hopf algebras. Hopf algebras are commonly called `quantum groups' because they
occur as generalised symmetries in statistical mechanics and also in some
quantum systems. The most famous quantum groups such as $U_q(g)$ as also
related to Poisson structures etc\cite{Dri} but they are {\em not} quantum
algebras of observables in a physical sense. In contrast to this well-known
line of development, there is in the literature a second independent origin or
quantum groups or Hopf algebras where they really do arise as quantum algebras
of observables. These are the {\em bicrossproduct Hopf
algebras}\cite{Ma:the}\cite{Ma:pla}\cite{Ma:phy}\cite{Ma:hop} and we shall be
concerned with these and recent work in the same line of development.

The physics behind these bicrossproduct Hopf algebras is that of quantum
mechanics on homogeneous spaces. These are a class of curved spaces in which
the quantisation scheme is perfectly clear and unambiguous. Namely, one can
quantise a particle on a homogeneous space in a standard way by making a
semidirect or cross product $\C(M)\lcross U(g)$ of the momentum group or Lie
algebra $g$ acting on the position observables $\C(M)$. Here $M$ is a manifold
on which $g$ acts, from the right say, and the particle is constrained to lie
on the orbits or homogeneous spaces for this action. The quantisation is
characterised by the commutation relations
\eqn{quantLie}{{} [\widehat{\xi},\widehat{f}]=\widehat{\widetilde\xi
(f)},\qquad \widetilde{\xi}(f)(s)={d\over dt}|_0 f(s\ra e^{t\xi}),\qquad s\in
M,\ \xi\in g, \ f\in \C(M)}
where $\widetilde\xi$ is the left-invariant vector field generated by $\xi$.
This is nothing more than Heisenberg's commutation relations in a co-ordinate
invariant form. All of this is quite standard. Sometimes, it also pays to work
with the group $G$ and make the semidirect product $\C(M)\lcross \C G$ by the
group algebra $\C G$ as a $C^*$-algebra or von Neumann algebra. At the
algebraic level this is
\eqn{quantgp}{ \widehat{u}\widehat{f}\widehat{u^{-1}}=\widehat{u\la f},\quad
(u\la f)(s)=f(s\ra u),\quad u\in G,\ f\in \C(M)}
where the given right action of $G$ on $M$ induces the left action $\la$ on the
position observables. Another version is to use Mackey's system of
imprimitivity, which is largely equivalent. Some early works on this topic are
\cite{Mac:ind}\cite{DobTol:mec}. In fact, the idea of cross products as
quantisation in this context is a conclusion reached independently by anyone
(including the author) thinking about this topic. If we do not worry too much
about the geometry then the cross product here is called a {\em dynamical
system}.

We begin in Section~2 by reviewing the results in \cite{Ma:the}--\cite{Ma:hop}
in this context, but in a simplified and more accessible algebraic form given
by working with finite groups or with Lie algebras. A reader who wants the full
functional-analysis for these models should see our original
paper\cite{Ma:hop}. The diagrammatic notion which we introduce is new.
Basically, we ask when do the quantisations on homogeneous spaces as above give
as a result of the quantisation a Hopf algebra? The physical implications of
this when it happens are quite deep and formed our algebraic approach to the
unification of quantum mechanics and geometry\cite{Ma:pla}, see also
\cite{Ma:ran}. The point is that not every homogeneous space or action $\ra$,
gives a Hopf algebra. The constraint even in one dimension forces the action to
be non-linear and the resulting motion far from rectilinear motion. In fact,
the resulting motion has some similarities with black-holes and we called the
constraint on the metric arising in this way a `toy version of Einstein's
equation'\cite{Ma:pla}\cite{Ma:the}. On the other hand, the famous quantum
double\cite{Dri} $D(G)$ on a group $G$ also  solves the condition and provides
a more trivial example of one of these bicrossproduct Hopf algebras. In this
case the orbits on which the particle moves are the conjugacy classes in $G$.

Next, we observe that with a little care the role of momentum group or Lie
algebra can be quite easily played by a Hopf algebra or quantum group $H$. Some
concrete examples of this are known, such as \cite{Egu:mec} for a particle on
a $q$-deformed 2-sphere, and \cite{Ma:mec} where examples are once again
provided by the quantum double, now in the general case $D(H)$. In particular,
$D(U_q(su_2))$, which is also called the {\em quantum Lorentz group} appears
now in a completely different light as the quantum algebra of observables of a
particle moving on a mass-shell in $q$-Minkowski space. This is a recent result
of the author\cite{Ma:mec} and we recall it breifly in Section~3. There are
many interesting features and physical questions raised by $q$-deforming the
above quantisations in this way. They are still quantum systems in so far as
they form $*$-algebras, but ones for which the underlying classical geometry is
a braided or $q$-deformed one.

In Section~4 we describe the extension theory of algebras by Hopf algebras.
This technology includes such famous things as the central extension of the
diffeomorphisms on a circle (the Virasoro algebra) but also includes and
generalises, with cocycles, all of the above cross products and
bicrossproducts. The cocycles here can be considered modulo an equivalence and
their classes form the nonAbelian cohomology spaces $\CH^2(H,A)$. An
interesting point is that these Hopf algebras unify some quite different
mathematical ideas into a single concept. Correspondingly different
constructions in physics are likewise of the same type from this general point
of view. A general theme in this area, which is also a theme of the paper, is
that this unification is quite typically between constructions that are usually
thought of as quantum mechanical  and constructions that are usually thought of
as geometric. For the author, this is the most important reason to study
quantum groups or Hopf algebras.

In Section~5 we explore this duality principle by pointing out that these same
cross products, which above are the quantum algebras of observables of quantum
systems, can also be thought of in terms of non-commutative geometry as trivial
quantum principal bundles. We use the general scheme of quantum group gauge
theory introduced in \cite{BrzMa:gau}. In this context the cohomology
$\CH^2(H,A)$ is a new kind of quantum number that exists even though the bundle
is trivial from the usual geometrical point of view. So they do not correspond
to any known geometrical cohomology of the base manifold etc, being a novel and
purely quantum possibility.

Finally, in Section~6 we show that a special case of the cohomology,
$\CH^2(H,\C)$, describes exactly the kind of cocycles that are needed in a dual
form of Drinfeld's theory of twisting\cite{Dri:qua} if one wants to twist a
Hopf algebra and remain as a Hopf algebra. This twisting theory gives an
alternative cohomological way to go about quantisation, as we shall demonstrate
on a formal example. This connection between Drinfeld's twisting and nonAbelian
cohomology is the mathematically new aspect of the present paper. Some of the
material will also be described in rather more detail in my forthcoming
book\cite{Ma:book}. The present paper nevertheless provides a self-contained
introduction for mathematical physicists to the topic.

\section{Bicrossproduct Hopf algebras revisited}

The Hopf algebra version of cross product quantisation is quite
straightforward, and we recall it first. Thus, a Hopf algebra is an algebra $H$
equipped with a coproduct $\Delta:H\to H\tens H$, a counit $\eps:H\to \C$ and
an antipode $S:H\to H$ obeying various axioms. The main ones are that $\Delta$
is an algebra homomorphism and is coassociative in the same sense as the
product is associative, but with arrows reversed. The antipode is like an
inverse. If $G$ is a finite group then $H=\C G$ the algebra generated by $1$
and the elements of the group (the group algebra) is a Hopf algebra with
$\Delta u=u\tens u$, $\eps u=1$ and $Su=u^{-1}$ for all $u\in G$. The most
unfamiliar thing about working with general Hopf algebras is the notation
$\Delta h=\nosum h\o\tens h\t$ for the explicit pieces of the tensor product
element -- a summation convention is to be understood here as there could be
more than one term in the description of such tensor product elements. There
are also some conventions associated with this to the effect that multiple
subscripts here can be renumbered to the canonical form with counting in base
ten. We refer to \cite{Swe:hop} for details. See also \cite{Ma:qua} for a
general introduction to quantum groups. Over $\C$ it is also natural to ask
that $H$ is a $*$-algebra, that $\Delta$ respects this and that $(S\circ
*)^2=\id$ as in \cite{Wor:twi}.

In this language it is easy enough to write down what is meant mathematically
by a Hopf algebra cross product quantisation. We have a Hopf $*$-algebra $H$ as
a generalised symmetry and a $*$-algebra $A$ to be viewed as like $\C(M)$, i.e.
we require that $H$ acts on $A$, and preserves its product and $*$ in the form
\eqn{modalg}{ h\la(ab)=\nosum (h\o\la a)(h\t\la b),\quad h\la 1=\eps(h),\quad
(h\la a)^*=(Sh)^*\la a^*.}
Given this, it is an easy exercise to to see that we have a {\em cross product
$*$-algebra} $A\lcross H$ built on the vector space $A\tens H$ with product and
$*$-structure
\eqn{crossprod}{ (a\tens h)(b\tens g)=\nosum a(h\o\la b)\tens h\t g,\quad
(a\tens h)^*=(1\tens h^*)(a^*\tens 1).}
This solves the quantisation problem as above by defining $\widehat h=1\tens h$
and $\widehat a=a\tens 1$, for then in the cross product we have
\eqn{quanthopf}{ \nosum \widehat{h\o}\widehat a\widehat{Sh\t}=\widehat{h\la a}}
which is clearly the correct generalisation of the familiar Lie group or Lie
algebra cases (\ref{quantLie})--(\ref{quantgp}) above.
Thus we have formulated the quantisation of particles on orbits in algebraic
terms.

Let us now ask abstractly when this algebra $A\lcross H$ is a Hopf algebra. Our
construction for these is based on the idea of keeping the input-output
symmetry between the algebra and the coalgebra, i.e. we introduce a coaction
$\beta$ and use it to twist the coproduct. The axioms of a coaction are like
those of an action but with all the maps reversed.

\begin{theorem}\cite{Ma:phy} Given a cross product as above, suppose further
that $A$ is a Hopf algebra and that there is a coaction $\beta:H\to A\tens H$
of $A$ back on $H$, which we write explicitly as $\beta(h)=\nosum h\bo\tens
h\bt$. If we have the {\em compatibility conditions}
\ceqn{bic}{\eps(h\la a)=\eps(h)\eps(a),\  \Delta(h\la a)=\nosum h\o\bo\la
a\o\tens h\o\bt(h\t\la a\t)\\
\beta(1)=1\tens 1,\ \   \beta(gh)=\nosum g\o\bo h\bo\tens g\o\bt ( g\t\la
h\bt)\\
\nosum h\t\bo\tens (h\o\la a)  h\t\bt =\nosum h\o\bo\tens h\o\bt( h\t\la  a) }
for all $a,b\in A$ and $h,g\in H$, then the cross product is a Hopf algebra,
the {\em bicrossproduct} $A\bicross H$, with coproduct, counit and antipode
\cmath{\Delta (a\tens h)=\nosum a\o\tens h\o\bo\tens a\t h\o\bt\tens h\t,\quad
\eps(a\tens h)=\eps(a)\eps(h)\\
 S(a\tens h)=\nosum (1\tens Sh\bo)(S(ah\bt)\tens 1)}
\end{theorem}

We content ourselves here with some concrete examples based on group
factorisations. Also, we will emphasise the structure for the finite case --
there is no problem for these with $*$-structures and representation on Hilbert
spaces needed for the quantum mechanical picture. Note that the question of
group or Lie algebra factorisations plays an important role in many areas of
mathematics and also in inverse scattering where the Lie case is sometimes
called a Manin-triple. On the other hand, the key idea for us is to express the
necessary data in an equivalent form as a pair of groups acting on each other.
So, two groups $(G,M)$ form a {\em matched pair} if there is a right action
$\ra$ of $G$ on $M$ and a left action $\la$ of $M$ on $G$ which are almost
actions on each other by automorphisms. The full set of conditions on the maps
are
\ceqn{mpair}{ s\ra e=s,\ (s\ra u)\ra v=s\ra(uv);\quad e\ra u=e,\ (st)\ra
u=\left(s\ra(t\la u)\right)(t\ra u)\\
 e\la u=u,\ s\la(t\la u)=(st)\la u;\quad s\la e=e,\ s\la(uv)=(s\la
u)\left((s\ra u)\la v\right)}
where the first two in each line says that we have an action, and the second
says that the action is almost by automorphisms, but twisted by the other
action. It is also useful to employ a graphical notation which expresses these
matching conditions as the ability to sub-divide rectangles. Thus we adopt the
notation that a square $s{\ \atop{\lform\atop \displaystyle u}}$ has on its top
boundary $u$ transformed by the action of $s$ and on its right boundary $s$
transformed by the action of $u$. Thus $s{\ \atop{\lform\atop \displaystyle
u}}=s{\displaystyle s\la u \atop { \lform\atop \displaystyle u}}s\ra u$. In
fact one can see that labelling any two adjacent edges is enough to uniquely
determine the other two edges to conform to this convention. We also adopt the
convention that any group elements labelling the same edge are to be read as
multiplied in the usual way for horizontal edges and downwards for vertical
edges. In this notation the matched pair conditions  become
\[\epsfbox{mpair.eps}\]
On the top right the condition is that a box with edges $st$ and $u$ can be
equally well viewed as a product of two boxes, one with edges $t$ and $u$ and
the other with one edge $s$ and the other edge the internal one labelled $t\la
u$. That the top edges agree on the two sides encodes the information that
$\la$ is an action, and that the vertical edges agree (when multiplied going
downwards) says the last condition in the first line of the matched pair
conditions (\ref{mpair}). The condition concerning the identity element $e$
says that a box with edge labelled by $e$ can be collapsed to one of zero
height, which notation is consistent with the gluing property. Likewise for the
second line.
This diagrammatic notation is borrowed from \cite{Mac:dou} where these notions
were generalised to groupoids. We use it now for the original bicrossproduct
Hopf algebras associated to matched pairs of groups as found in \cite{Ma:phy}
as a case of Theorem~2.1. We use $\ra$ to make a cross product and $\la$ to
define the coaction $\beta(u)=\sum_s s\la u\tens\delta_s$ for a cross
coproduct. It turned out that this special case was also known from
\cite{Tak:mat} in another context.

\begin{propos} If $(G,M)$ is a matched pair of groups then the induced action
of $\C G$ on $\C(M)$ and coaction of $\C(M)$ on $\C G$ gives a left-right
bicrossproduct Hopf algebra $\C(M)\bicross \C G$ as in Theorem~2.1. In basis
$\{\delta_s\tens u\}$ the structure is explicitly
\cmath{ (\delta_s\tens u)(\delta_t\tens v)=\delta_{{s\ra u},t}(\delta_s\tens
uv),\quad \Delta(\delta_s\tens u)=\sum_{ab=s}\delta_a\tens b\la u\tens
\delta_b\tens u\\
1=\sum_s\delta_s\tens e,\quad \eps(\delta_s\tens u)=\delta_{s,e},\quad
S(\delta_s\tens u)=\delta_{(s\ra u)^{-1}}\tens(s\la u)^{-1}.}
In the graphical notation this is
\[ \epsfbox{bichopf.eps}\]
\end{propos}
\proof Here we want to provide a new and direct diagrammatic verification of
the Hopf algebra axioms. Recall from our explanation above that two adjacent
edges of a box determine the other two. We add to this the convention that two
boxes can be multiplied by gluing as shown if the vertical edges match in their
values. Otherwise the product is zero. This is the product in $\C(M)\bicross \C
G$.  We use the other dimension in a similar but dual way to make the
coproduct. Thus the coproduct of a box is the sum over labelled boxes such that
when glued vertically they would give the labelled box we began with. Our
convention is to read vertical expressions from top to bottom, so the upper box
is the first output of the coproduct and the lower box is the second. This
explains the diagrammatic form of the Hopf algebra structure. The proof that
$\Delta$ is an algebra homomorphism is given in this language by
\[ \epsfbox{bicproof.eps}\]
We compute $\Delta$ on a composite in the first line. The third equality
decomposes each of the blocks of the coproduct into pieces. The subdivision
picture of the matched pair conditions tells us exactly that this can be done
in a way that the internal  parallel edges in the fourth expression have
matching values as needed for gluing. All the edges of all four boxes are full
determined by $a,b,u,v$ according to the above conventions.  In particular, the
values $c,d$ for the left edges of the right-hand boxes must have product
$cd=(ab)\ra u$ which is the product of the right edges of the left-hand boxes.
This is shown in the fifth expression. But written in this way, we have the
product pairwise in two copies of the the Hopf algebra of the outputs of
$\Delta$ as required. Similarly for the other structures. \endproof

The matched pair obviously has a left-right symmetry, so not surprisingly there
is another bicrossproduct Hopf algebra  $\C M\cobicross \C(G)$ constructed in a
similar way but with $\la$ supplying the action and $\ra$ the coaction. It is
built on $\C M\tens \C(G)$ with basis $\{s\tens \delta_u\}$, say, and structure
\cmath{(s\tens \delta_u)(t\tens \delta_v)=\delta_{u,t\la v}(st\tens
\delta_{uv}),\quad \Delta (s\tens \delta_u)=\sum_{vw=u}s\tens\delta_v\tens s\ra
v\tens \delta_w\\
 1=\sum_u e\tens \delta_u,\quad \eps (s\tens \delta_u)=\delta_{u,e},\quad
S(s\tens \delta_u)=(s\ra u)^{-1}\tens\delta_{(s\la u)^{-1}}}
\[ \epsfbox{cobichopf.eps}\]
One has that it is the dual Hopf algebra to the previous one by the evaluation
pairing,
\eqn{dualbic}{\C M\cobicross \C(G)=(\C(M)\bicross \C G)^*.}

One can also construct from our matched pair a {\em double cross product group}
$G\dcross M$ defined as the set $G\times M$ with product and inverse
\cmath{ (u,s)(v,t)=(u(s\la v),(s\ra v)t),\quad (u,s)^{-1}=(s^{-1}\la
u^{-1},s^{-1}\ra u^{-1})}
\[\epsfbox{dgroup.eps}\]
If a group $X$ factorises in the sense that $G{\buildrel
i\over\hookrightarrow}X{\buildrel j\over\hookleftarrow}M$ are two subgroups and
the map $G\times M\to X$ given by multiplication in $X$ is a bijection, then
$(G,M)$ are a matched pair and $X\isom G\dcross M$. The actions are determined
by
\[j(s)i(u)=i(s\la u)j(s\ra u).\]
This is why the matched pair data are in one-to-one correspondence with group
factorisations. As such, they are abundant in nature. One can see \cite{Ma:phy}
for some relevant examples according to our current physical interpretation. We
content ourselves with the simplest one-dimensional one of these.
So let $G=M=(\R,+)$ with its additive group structure. Then the general
solution of the matching conditions (\ref{mpair}) in a neighbourhood of the
origin has two parameters $\sA,\sB\in \R$ and the
form\cite{Ma:phy}\cite{Ma:pla}
\eqn{mpairR}{s\ra u={1\over \sB}\ln(1+(e^{\sB s}-1)e^{-\sA u}),\quad s\la
u={1\over \sA}\ln(1+e^{-\sB s}(e^{\sA u}-1)).}
Plugging these into the $C^*$ algebra or Hopf-von Neumann algebra version of
Proposition~2.2 gives the quantum algebra of observables for this
system\cite{Ma:pla}. Note that the general Hopf algebra version in Theorem~2.1
applies just as well to enveloping algebras as to group algebras, and at this
level the same solution (\ref{mpairR}) gives gives the following bicrossproduct
Hopf algebra.

\begin{example}\cite{Ma:pla} We take $A=\C[g,g^{-1}]$ and $H=U(\R)=\C[p]$.
These form a bicrossproduct with action and coaction
\[ p\la g=\sA(1-g)g,  \quad \beta(p)=p\tens g.\]
The cross product algebra and cross coproduct coalgebra are
\cmath{ [  p ,g]=\sA g(1-g),\quad \Delta g=g\tens g\\
\Delta  p=  p\tens g+1\tens p,\ \eps g=1,\ \eps p=-p,\quad Sg=g^{-1},\quad
Sp=-pg^{-1}}
where we omit writing the quantisation maps $\widehat{\ }$.
\end{example}
Note that if we work formally over $\C[[B]]$ with $x$ as the generator and
$g=e^{-\sB x}$, and if we set $\hbar=-{\sA\over \sB}$ then
\[ p\la x=\hbar(1-e^{-\sB x}),\quad [  p,  x]=\hbar(1-e^{-\sB  x})\]
is the cross product algebra. The stated action and coaction here are obtained
by computing the action and coaction at the group level for the group actions
(\ref{mpairR}) and then differentiating with respect to the group $G=\R$. This
is because we are replacing it by the enveloping algebra $U(\R)=\C[p]$. We keep
our position space $M=\R$ in the additive form but work with its co-ordinate
function $x(s)=s$ or, more precisely, with the function $s\mapsto e^{-\sB s}$
as the abstract generator $g$.
After obtaining these formulas one then verifies them directly in our algebraic
picture. Thus one can show that they extend uniquely to an action and coaction
fulfilling the conditions of Theorem~2.1 for a bicrossproduct.

This Hopf algebra $\C[g,g^{-1}]\bicross \C[p]$ or $\C[x]\bicross\C[p]$ is one
of the simplest non-commutative and non-cocommutative Hopf algebras, and was
introduced along the lines above by the author. We now discuss in detail its
physical meaning from \cite{Ma:pla}. Firstly, it is significant that this
general class of constructions based on keeping the group structure of phase
space in the quantum domain, allows us only two free parameters $\sA,\sB$ with
action and reaction as stated above. All possible quantisations of a particle
in one dimension for which the quantum algebra of observables remains a Hopf
algebra of self-dual type are classified by two parameters. We have already
identified the combination $\hbar=-{\sA\over\sB}$ by looking at large $x\sB$.
To see the meaning of the remaining parameter $\sB$ we consider how the
particle moves classically. In our approach we keep $p$ as a conserved momentum
and keep Hamiltonian ${-p^2\over 2m}$ so that the particle is in free-fall. The
different motions of the particle are then controlled by changing the $x,p$
commutation relations. It is more usual to keep the commutation relations fixed
in a canonical form and vary the Hamiltonian, and indeed we could reformulate
things this way except that the necessary change of variables would have to be
singular. This is because the usual quantum mechanics algebra is not a Hopf
algebra while our bicrossproduct one, is. Note also that in our conventions $p$
is antihermitian, i.e. $-\imath p$ is the physical momentum observable. Then
\[ {dx\over dt}={\imath\over\hbar}[-{p^2\over 2m},x]=({-\imath p\over
m})(1-e^{-Bx})+ O(\hbar),\quad {dp\over dt}=0\]
as operators. Hence in representations where the system behaves like a
particle, its classical trajectories will be of the form given here by the
leading term. We identify ${-\imath p\over m}=v_{\infty}$, the velocity at
$x=\infty$. If we consider a particle falling in from $\infty$ then we see that
the particle approaches the origin $x=0$ but does so more and more slowly. In
fact, it takes an infinite amount of time to reach the origin, which therefore
behaves in some ways like a black-hole event horizon. The present model is one
dimensional but we can imagine that it is the radial part of some motion in
spacetime, and can estimate the value of $\sB$ on this basis. We find
\[ \sB={c^2\over \sM \grav};\quad {d x\over dt}=v_\infty\left(1-{1\over
\exp({c^2 x\over \sM \grav})}\right),\quad {\rm cf.}\  {dx\over
dt}=-c\left(1-{1\over 1+\h{c^2x\over \sM \grav}}\right)\]
where the comparison is with an in-falling photon at radial distance $x$ from
the event horizon in the Schwarzschild black-hole solution of mass $\sM $. Here
$c$ is the speed of light and $\grav$ the gravitational coupling constant. This
analogy should not be pushed too far since our present treatment is in
non-relativistic quantum mechanics, but it gives us at least one interpretation
of the parameter $\sB$ as being comparable to introducing the distortion in the
geometry due to a gravitational mass $\sM $.

Another, more mathematical way to reach the same conclusion is to take the
limit $\hbar\to 0$. In this case our algebra becomes commutative, but the
coalgebra remains non-cocommutative. In this case $\C[x]\bicross \C[p]\isom
\C(X )$ where $X $ is the group $\R\rcross\R$ with group law
\[ X =\{(s,u)\},\quad (s,u)(t,v)=(s+t,ue^{-\sB t}+v)\]
To see this, let $p(s,u)=u$ and $x(s,u)=s$ be the co-ordinate functions on $X
$. They commute, and with coproducts determined from
\cmath{(\Delta x)((s,u),(t,v)) =x((s,u)(t,v))=s+t=(x\tens 1+1\tens
x)((s,u),(t,v))\\
(\Delta p)((s,u),(t,v)) =p((s,u)(t,v))=ue^{-\sB t}+v=(p\tens e^{-\sB x}+1\tens
p)((s,u),(t,v))}
they generate our algebraic model $\C(X )$ of the functions on $X $. This is
clearly the limit of $\C[x]\bicross \C[p]$.    This group $\R\rcross\R$ is
therefore the underlying classical phase space of our system. The coproduct
equips it with a nonAbelian group structure. In geometrical terms a nonAbelian
group law corresponds to geometrical curvature. Note that since the group is
not semisimple (it is solvable), its natural metric induced by the action is
degenerate. So there are some subtleties here but the general principle is the
same. In our case one can compute that the curvature on phase space is of the
order of $\sB^2$. For dynamical models with a reasonable degree of symmetry one
can expect that this should also be comparable to the curvature in position
space. Comparing it with the curvature near a mass $\sM $ for example would
give the same estimate as above.

In summary, we see that the bicrossproduct model has two limits:

\begin{picture}(300,50)(-10,-15)
\put(0,0){$ \C[x]\bicross
\C[p]$}\put(55,8){\vector(3,1){28}}\put(55,0){\vector(3,-1){28}}
\put(86,13){$\C[x]\lcross \C[p]$ usual quantum mechanics for $x>0$}
\put(86,-13){$\C(X )$ usual curved geometry}
\put(36,15){\sevrm $\grav\to 0$}\put(36,-15){\sevrm $\hbar\to 0$}
\end{picture}

\medskip
\noindent This illustrates the first goal of \cite{Ma:pla} of unifying a
quantum system and a geometrical one. The most general unification within this
framework allows only two free parameters, which we have identified in general
terms as $\hbar,\grav$. Moreover, the existence of curvature in our geometrical
system forces the dynamics of the quantum particle to be deformed from the
usual one. This deformation forces, as we have seen, a structure not unlike a
black-hole event horizon. With a little more care one can estimate also how
small $\hbar,\grav$ have to be in comparison to the other scales in the system.
For example, we can take the two scales in the system as $m$, the mass of the
quantum test-particle moving in the curved background and $\sM $, the active
gravitational mass which we estimate as causing comparable curvature. The
flat-space quantum-mechanical picture is valid if $\sB x\gg 1$.  If we consider
a relativistic quantum particle then its position is only defined up to its
Compton wavelength $\hbar\over mc$  so that the condition that the system is
not detectably different from usual flat-space quantum mechanics in the region
$x>0$ is
\[ m\sM \ll m_{\rm Planck}^2.\]
On the other hand, the system appears classical if $\hbar\ll \Delta p\Delta x$.
If we estimate $\Delta p\sim mc$ again and suppose that the length scale of
interest obeys a natural curvature constraint $\Delta x\gg {1\over \sB}$ coming
out of general relativity, then the condition that the quantum aspect of the
algebra is not detectable appears as
\[ m\sM \gg m_{\rm Planck}^2.\]

Some of the general features here also hold for other bicrossproduct models
associated to group factorisations. The action $\ra$ describes in general the
flow or `metric' under which the quantum particle moves. But not every action
admits a back-reaction $\la$ forming a matched pair in the sense above. This is
a genuine constraint, which one can think of as an integrated form of a second
order differential equation for the action. Moreover, this constraint in the
above examples, and also for more complicated examples in\cite{Ma:mat}, does
have qualitative similarities with the singularities forced by Einstein's
equation for the metric in the presence of matter. From this point of view, the
matched pair conditions (\ref{mpair}) are some kind of toy version of
Einstein's equations. For more complicated metrics one would need of course to
leave the class of Hopf algebras and construct more complicated quantum
geometries, but exhibiting perhaps some of these same features.

Next we note that by their construction, the bicrossproduct models are
self-dual as we know from the symmetry in the matched pair $\la,\ra$. To see
this more precisely one has to introduce some functional analysis and work with
Hopf-von Neumann algebras, or else proceed with formal power-series. For our
algebraic purposes we construct directly the corresponding Hopf algebra
$\C[\phi]\cobicross \C[\psi]$ say, and show that it is dually paired.
The construction follows this time the right action and left coaction version
of  Theorem~2.1 as,
\cmath{ \psi\ra \phi={\hbar^{-1}}(1-e^{-\sA\psi}),\quad
\beta(\phi)=e^{-\sA\psi}\tens\phi,\quad [\psi,\phi]=\hbar^{-1}(1-e^{-\sA\psi})
\\
\Delta \phi=\phi\tens 1+e^{-\sA\psi}\tens\phi, \quad \Delta \psi=\psi\tens
1+1\tens\psi\\
 \eps\phi=\eps\psi=0,\quad S\phi=-e^{A\psi}\phi,\quad S\psi=-\psi}
We have followed here exactly the same steps as in Example~2.3 but with the
roles of the two groups or the roles of $\sA,\sB$, interchanged. Next, we
should think of $\C[\phi]$ as the universal enveloping algebra $U(\R)$ and
hence dually paired with $\C[x]$. This pairing is
\[ \<\phi^n,x^m\>=\delta^n{}_m n!,\quad {\rm i.e.},\quad \<
\phi^n,f(x)\>={d^n\over dx^n}|0 f.\]
In the same way, $\C[p]$ is dually paired with $\C[\psi]$. Along the same lines
as (\ref{dualbic}), we conclude that $\C[\phi]\cobicross \C[\psi]$ is dually
paired with $\C[x]\bicross \C[p]$. Explicitly, it is
\[ \<\phi, :f(x,p):\>=({\del f\over \del x} )(0,0),\quad
\<\psi,:f(x,p):\>=({\del f\over \del p} )(0,0)\]
where $:f(x,p):=\sum f_{n,m} x^np^m$ is the normal-ordered form of a function
in the two variables $x,p$. For a purely polynomial version, one should work
with $e^{-Bx}$ and $e^{-A\psi}$ as the generators as explained above.

Now we note that linear functionals on a quantum Hopf algebra of observables,
form themselves an `algebra of states'. The physical states are the positive
ones among them\cite{Ma:pla}. In our model then, the algebra of states is
generated by the linear functionals $\phi,\psi$. Moreover, we see that this
algebra of states is exactly like a quantum system. It is natural to give
$\phi$ the dimensions of inverse length, and $\psi$ the dimensions of inverse
momentum. Moreover, if we consider
\[ x'=\hbar\psi,\quad p'=\hbar e^{A\psi}\phi\]
we see that these have dimensions of length and momentum and have exactly the
same Hopf algebra structure as $\C[x]\bicross \C[p]$. Thus one could equally
well regard this second Hopf algebra as the algebra of observables, with $p'$
as momentum and $x'$ as position. One would have the same picture as above in
terms of geometry and quantum mechanics. The possibility to make such a
reinterpretation of the same algebraic structures is the second and more
radical theme that the bicrossproduct models demonstrate\cite{Ma:pla}. In this
reinterpretation, the roles of observables and states become reversed. Hence
also the roles of non-commutativity in the algebra (of quantum origin) and
non-cocommutativity in the coalgebra  (of geometrical origin as curvature on
phase space), become reversed. This is a new kind of symmetry principle which
one can propose as a speculative idea for the structure of Planck scale
physics. This is the novel physical phenomenon underlying the bicrossproduct
Hopf algebras. We refer to \cite{Ma:the}\cite{Ma:pla}\cite{Ma:pri}\cite{Ma:the}
for further details and discussion of its meaning.

\section{Example of the quantum double}

A simple case of Proposition~2.2 is when the back-reaction $\la$ is trivial. In
this case the conditions of a matched pair become just that the remaining
action of $G$ on $M$ is by group automorphisms. This is just what it takes then
for $\C(M)\lcross \C G$ to be a Hopf algebra with tensor product coalgebra
structure.

A simple example of this type is when $G$ acts on $M=G$ by the right adjoint
action. So
\ceqn{DG}{s\ra u=u^{-1} s u,\quad s\la u=u,\qquad u,s\in G\\
 (\delta_s\tens u)(\delta_t\tens v)=\delta_{u^{-1}su,t}(\delta_s\tens uv),\quad
\Delta(\delta_s\tens u)=\sum_{ab=s}\delta_a\tens u\tens\delta_b\tens v\\
\eps(\delta_s\tens u)=\delta_{s,e},\quad S(\delta_s\tens
u)=\delta_{u^{-1}s^{-1}u}\tens u^{-1}}
which is the nowadays called the {\em quantum double} $D(G)=\C(G){}_{{\rm
Ad}^*}\lcross\C G$ of the group $G$. According to our picture in \cite{Ma:the}
etc, as above, one can think of it as the quantum algebra of observables of a
particle moving on conjugacy classes in $G$.

Drinfeld in \cite{Dri} introduced a more general construction of a quantum
group $D(H)$ for any finite-dimensional Hopf algebra $H$. We do not want to
recall it in detail -- it is not manifestly a cross product of the type above,
but rather an example of {\em double cross product} construction as introduced
in \cite{Ma:phy} in a  theory of Hopf algebra factorisations. Nevertheless, we
have shown in \cite{Ma:skl}\cite{Ma:mec} that if $H$ is a real-quasitriangular
Hopf $*$-algebra then the double {\em is} isomorphic to a $*$-algebra cross
product
\[ D(H)\isom B\lcross H\]
where $B$ is a {\em braided group} of function algebra type\cite{Ma:bg}. On the
other hand, if $H$ is anti-real quasitriangular then it is the {\em dual} of
the quantum double which is isomorphic to a $*$-algebra cross product based on
the dual of $B$ and the dual of $H$. The two cases are
\[ \CR^{*\tens *}=\CR_{21},\quad {\rm (real)},\qquad \CR^{*\tens
*}=\CR^{-1},\quad {\rm (anti-real)}\]
where $\CR$ is the quasitriangular structure or universal R-matrix\cite{Dri}.
In one case we have that the quantum double is some quantum system in that it
forms a $*$-algebra cross product; in the other case the dual of the quantum
double is some other quantum system in this sense.

Thus we are in the general situation with observable-state symmetry as above.
One big difference of course is that the generalised momentum group $H$ and the
position observables $B$, etc, are now quantum or braided groups. But this does
not prevent us from having an honest quantum system in the sense of a
$C^*$-algebra or von Neumann algebra, after suitable functional analysis. It
just means that we have a quantum system whose natural underlying `classical
system' is not usual geometry. In the example given in detail in \cite{Ma:mec},
the classical system is rather a $q$-deformed geometry. Our point of view
elavates the idea of an abstract quantum system expressed via an algebra of
observables and states to first place. It is the starting point which may or
may not have a familiar or recognisable classical limit. Too often in physics
one begins with the classical system as something given which we must quantise,
a view which we consider strictly unnecessary.

We content ourselves here with outlining the concrete example based on
$q$-Minkowski space and $q$-rotations in \cite{Ma:mec},  to which we refer the
reader for details. Briefly, for our position observables we begin with the
$*$-algebra $BH_q(2)$ of $2\times 2$ braided-hermitian matrices
\[ \pmatrix{a^* & b^*\cr c^*
&d^*}=\pmatrix{a&c\cr b&d}\]
The co-ordinate functions here do not commute but rather they form the algebra
of $2\times 2$ braided matrices $BM_q(2)$ introduced in \cite{Ma:exa}
\[\nqquad qd+q^{-1}a{\rm\ central},\  ba=q^2ab,\ ac=q^2ca,\
bc=cb+(1-q^{-2})a(d-a);\ q\in\R.\]
One can write these relations in the compact form
$R_{21}\vecu_1R\vecu_2=\vecu_2 R_{21}\vecu_1 R$ which is nowadays popular in a
number of contexts, as explained in \cite{Ma:skl}. For the braided matrices in
\cite{Ma:exa} these relations were introduced precisely as {\em braided
commutativity} relations. This is even clearer if one uses the multi-index
notation of \cite{Ma:exa} where $\{u^i{}_j\}$ are regarded as a 4-vector. Then
the relations can be written in the quantum-plane form $uu= uu \Psi'$ with
suitable indices and with $\Psi'$ a solution of the quantum Yang-Baxter
equations built from four copies of $R$\cite{Ma:exa}.  From this point of view,
$q$-Minkowski space is like a super-space in so far as its co-ordinates are
commutative after allowing for some (braid) statistics. The $q$ here has to do
with the statistics of our classical system and not to do with quantisation.

One can consider a particle moving on this $q$-deformed algebra of position
observables. The $q$-Lorentz group in the form of a certain Hopf $*$-algebra
(related in fact to the quantum double of $U_q(su_2)$ as symmetry quantum
group) acts on it. In \cite{Ma:mec} we specialised further to the Lorentzian
sphere $BS_q^2$ in $q$-Minkowski space by setting $BDET(\vecu)\equiv ad-q^2
cb=1$. This is preserved by the $q$-rotation subgroup of the $q$-Lorentz group,
of which the double cover is $U_q(su_2)$. So this acts on the algebra of
functions on our $q$-Lorentzian sphere as angular momentum. The explicit action
was given in \cite{Ma:exa} and is based on the quantum adjoint action. We
consider then a particle moving on $A=BS_q^2$ as the position observables of
the system and $H=U_q(su_2)$ as generalised momentum group (so a kind of
$q$-deformed Lorentzian top). We then quantise the system according to the
generalised Mackey scheme above and obtain a $*$-algebra cross product
isomorphic to the quantum double. The overall situation is
\[ \matrix{ \C(S^3_{\rm Lor})\tens su_2&{\buildrel {\rm
deformation}\over \to}& BS^3_q\tens U_q(su_2)\cr
& & \cr
{\rm \scriptstyle quantisation}\ \downarrow& &\downarrow{\rm \scriptstyle \
quantisation}\cr
& & \cr
\C(S^3_{\rm Lor})\lcross U(su_2)&{\buildrel {\rm deformation}\over
\to}&BS^3_q\lcross U_q(su_2).}\]
Note that at $q=1$ we could just as well take the cross product
$\C(SU(2))\lcross U(su_2)$, which is a particle moving on a Euclidean 3-sphere
and which is a quantum double of the type (\ref{DG}) with the finite group
replaced by $SU(2)$. But this picture does {\em not} extend to the $q$-deformed
case as a $*$-algebra cross product because of a breakdown of covariance for
quantum matrices. Instead, to keep covariance one must use the braided
matrices\cite{Ma:exa}, which then forces the Lorentzian signature. In summary,
which is the important point for physics:
\begin{quote}
{\em the possibility of $q$-deformation in this context forces the Lorentzian
signature of spacetime}
\end{quote}
There is also a dual quantum system valid when $|q|=1$. It has position
$BSL_q(2,\R)$ and momentum quantum group $U_q(su_2^*)$\cite{Ma:mec}.

\section{Quantum anomalies and nonAbelian cohomology}

Now we are going to consider cocycle versions of some of the above
constructions. Cocycles of the type we consider are associated in physics with
quantum anomalies. Thus, a symmetry group or Lie algebra $g$ may not be
properly represented in our quantum system but rather only projectively
represented. Instead then, the extended group or Lie algebra is represented. We
recall that a Lie algebra extension means
\eqn{ext-lie}{
[\widehat\xi,\widehat\eta]=\widehat{[\xi,\eta]}+\widehat{\chi(\xi,\eta)}}
where the extension is denoted by $\widehat{\ }$ and $\chi$ is a Lie algebra
2-cocycle. At the group level this becomes
 \eqn{ext-gp}{ \widehat u\widehat v=\widehat{uv}\widehat{\chi(u,v)}}
for $\chi$ a group 2-cocycle. This could have values in $S^1$ or more generally
in any Abelian group $M$. Then the extension $M\hookrightarrow
E\twoheadrightarrow G$ in this case can be built as a new group on the set
$E=G\times M$ with product
\eqn{ext-gp-E}{ (u,s)(v,t)=(uv,\chi(u,v)st),\quad \widehat u=(u,e),\quad
\widehat s=(e,s).}
It is clear that $M$ is a subgroup in the center. Some famous examples are the
Virasoro algebra and loop group extensions.

Coming now to our original physical setting of cross products as quantisation,
we ask under what conditions cocycles $\chi$ might be observed. Recall that
$\C(M){}_\chi\lcross \C G$ is the quantisation of a system with $G$ in the role
of momentum group and $M$ as position space. The point is that $\C G$ does not
appear here as a subalgebra as soon as $\chi$ is present. Rather, we have
exactly the situation as above but as algebras and with the cocycle now having
values in the algebra $\C(M)$ rather than in a group. Recall that $M$ is the
position space. So this time
\eqn{cocyquant}{ \C(M)\hookrightarrow\C(M){}_\chi \lcross \C G,\quad \widehat
u\widehat f\widehat{u^{-1}}=\widehat{u\la f},\quad \widehat u\widehat
v=\widehat{uv}\widehat{\chi(u,v)}}
is the cocycle cross product quantisation of particles on a homogeneous space,
here at the group level. Similarly at the Lie algebra level. This is exactly
the situation which arises in some models under the heading of a quantum {\em
anomaly}. One does not often expect the momentum symmetry to be anomalous, but
something similar does happen for the diffeomorphism group in conformal field
theory, for example. Also, there is a standard notion of equivalence of
cocycles up to coboundary and hence of group cohomology $\CH(G,\C(M))$ which
controls the anomaly. The cohomologically trivial sector is the non-anomalous
one.

We proceed now to extend all these ideas to the Hopf algebra setting, just as
we did for ordinary cross products in Section~2. Apart from unifying the Lie
algebra and group cases into a single setting, some purely quantum
possibilities also arise as we shall see. We will end with the example of the
{\em quantum Weyl group} as explained in \cite{MaSoi:bic}.
The theory of general cocycle bicrossproduct Hopf algebras was introduced by
the author\cite{Ma:mor}. The cohomology shows up even at the algebra level,
where it was studied previously by Y. Doi\cite{Doi:equ}\cite{DoiTak:cle} and
subsequent authors. In fact, the main ideas here are already in earlier works
of Hyeneman and Sweedler and Chase and Sweedler for special cases. Likewise,
the cohomological picture at the bialgebra level was studied in the special
Abelian or (co)commutative case in \cite{Sin:ext}.

Consider then the same quantisation problem as in Section~2, but a bit more
generally. We have a Hopf algebra $H$ and an algebra $A$ and suppose that $A$
is `acted' upon by $H$ via a map $\la:H\tens A\to A$ such that the product and
unit are respected in the sense of (\ref{modalg}). We do not require $\la$ to
be an action in the usual sense but rather a {\em cocycle action}. This means
there is a map $\chi:H\tens H\to A$ such that
\ceqn{cy-lmod}{ \nosum h\o\la(g\o\la a)\chi(h\t\tens g\t)=\nosum\chi(h\o\tens
g\o)((h\t g\t)\la a),\quad 1\la a=a}
and moreover, this $\chi$ is a {\em 2-cocycle} in the sense
\ceqn{2-cy-mod}{\nosum \left(h\o\la \chi(g\o\tens f\o)\right)\chi(h\t\tens g\t
f\t)=\nosum \chi(h\o\tens  g\o)\chi(h\t g\t\tens f)\nonumber\\
\chi(1\tens h)=\chi(h\tens 1)=1\eps(h)}
for all $h,g,f\in H$ and $a,b\in A$. The cocycle terminology here is justified
by the following two propositions.

\begin{propos} Let $A$ be an algebra, $H$ a bialgebra and $(\chi,\la)$ a
cocycle-action as above. If $\gamma:H\to A$ is a convolution-invertible linear
map with $\gamma(1)=1$, then
\cmath{\chi^\gamma(h\tens g)=\nosum \gamma(h\o)(h\t\la
\gamma(g\o))\chi(h\th\tens g\t)\gamma^{-1}(h\fo g\th)\\
h\la^\gamma a=\gamma(h\o)(h\t\la a)\gamma^{-1}(h\th)}
is also cocycle-action. We say that $(\chi,\la)$ and $(\chi^\gamma,\la^\gamma)$
are {\em cohomologous} and denote the equivalence classes under such
transformations by $\CH^2(H,A)$.
\end{propos}
\proof This is known from previous work\cite{Doi:equ}, but is also easy enough
for us to prove directly.
It is a straight verification from the definition of a cocycle.
Convolution-invertible means that there is a map $\gamma^{-1}:H\to A$ such that
$\cdot\circ(\gamma\tens\gamma^{-1})\circ\Delta=\eps
=\cdot\circ(\gamma^{-1}\tens\gamma)\circ\Delta$. Coassociativity gives at once
that $\la^\gamma$ obeys (\ref{modalg}) if $\la$ does. Next we compute
\align{&&\equad h\o\la^\gamma (g\o\la^\gamma a)\chi^\gamma(h\t\tens g\t)\\
&&=\gamma(h\o)\left(h\t\la(\gamma(g\o)(g\t\la
a)\gamma^{-1}(g\th))\right)\gamma^{-1}(h\th)\\
&&\qquad\qquad \gamma(h\fo)(h\fiv\la \gamma(g\fo))\chi(h\six\tens
g\fiv)\gamma^{-1}(h\sev g\six)\\
&&=\gamma(h\o)(h\t\la \gamma(g\o)) (h\th\la (g\t\la a))( h\fo \la
(\gamma^{-1}(g\th)\gamma(g\fo)))\\
&&\qquad\qquad\chi(h\fiv\tens g\fiv)\gamma^{-1}(h\six g\six)\\
&&=\gamma(h\o)(h\t\la \gamma(g\o))\chi(h\th\tens g\t) ((h\fo g\th)\la
a)\gamma^{-1}(h\fiv g\fo)\\
&&=\gamma(h\o)(h\t\la \gamma(g\o))\chi(h\th\tens g\t) \gamma^{-1}(h\fo g\th)\\
&&\qquad\qquad \gamma(h\fiv g\fo) ((h\six g\fiv)\la a)\gamma^{-1}(h\sev
g\six)\\
&&=\chi^\gamma(h\o\tens g\o)((h\t g\t)\la^\gamma a)}
as required for (\ref{cy-lmod}). We used for the third equality the same
property for $\chi,\la$. Likewise, we verify
\align{&&\equad h\o\la^\gamma\chi^\gamma(g\o\tens f\o)\chi^\gamma(h\t\tens g\t
f\t)\\
&&=\gamma(h\o)( h\t\la\left(\gamma(g\o)(g\t\la \gamma(f\o))\chi(g\th\tens
f\t)\gamma^{-1}(g\fo f\th)\right))\gamma^{-1}(h\th)\\
&&\qquad\qquad \gamma(h\fo)(h\fiv\la \gamma(g\fiv f\fo))\chi(h\six\tens g\six
f\fiv)\gamma^{-1}(h\sev g\sev f\six)\\
&&=\gamma(h\o)(h\t\la \gamma(g\o))(h\th\la(g\t\la \gamma(f\o)))\\
&&\qquad\qquad (h\fo\la\chi(g\th\tens f\t))\chi(h\fiv\tens g\fo
f\th)\gamma^{-1}(h\six g\fiv f\fo)\\
&&=\gamma(h\o)(h\t\la\gamma(g\o))(h\th\la (g\t\la\gamma(f\o)))\\
&&\qquad\qquad \chi(h\fo\tens g\th)\chi(h\fiv g\fo\tens f\t)\gamma^{-1}(h\six
g\fiv f\th)\\
&&=\gamma(h\o)(h\t\la \gamma(g\o))\chi(h\th\tens g\t)\\
&&\qquad\qquad ((h\fo g\th)\la\gamma(f\o))\chi(h\fiv g\fo\tens
f\t)\gamma^{-1}(h\six g\fiv f\th)\\
&&=\chi^\gamma(h\o\tens g\o)\chi^\gamma(h\t g\t\tens f)}
as required. The second equality uses (\ref{modalg}) and makes cancellations of
$\gamma,\gamma^{-1}$ as before. The third uses the 2-cocycle condition
(\ref{2-cy-mod}) for $\chi$. The fourth equality uses (\ref{cy-lmod}) for
$\chi,\la$. The properties with respect to the unit are clear provided
$\gamma(1)=1$. It is also easy to see that this transformation by $\gamma$
gives a left action of the group of identity-preserving convolution-invertible
maps $H\to A$, i.e. $(\chi^\mu)^\gamma=\chi^{\gamma*\mu}$ and
$(\la^\mu)^\gamma=\la^{\gamma*\mu}$ where
$\gamma*\mu=(\gamma\tens\mu)\circ\Delta$ is the convolution product of two such
maps. \endproof

Armed with such cocycle-actions we can make a (cocycle) cross product algebra
$A{}_\chi\lcross H$. It is built on $A\tens H$ with product
\eqn{cocyclecross}{ (a\tens h)(b\tens g)=\nosum a(h\o\la b)\chi(h\t\tens
g\o)\tens h\th g\t}
and the unit element is $1\tens 1$. One can easily see that the cocycle
conditions correspond to this being associative. Note also that $H$ coacts on
itself from the right and hence also on $A{}_\chi\lcross H$ by the algebra
homomorphism $\beta=\id\tens \Delta$. We will say that two such cocycle cross
product algebras are {\em regularly isomorphic} or equivalent if they are
isomorphic by a map which is covariant under this coaction. We require also
that the isomorphism restricts to the identity on $A$ as a subalgebra of the
cross product.

\begin{propos}  Two cocycle cross products are regularly isomorphic iff their
corresponding cocycle module algebra structures are cohomologous, i.e. the
equivalence classes of cocycle cross products $A{}_\chi\lcross H$ are in
one-one correspondence with elements of the nonAbelian cohomology $\CH^2(H,A)$.
In particular, $A{}_\chi\lcross H$ is regularly isomorphic to the tensor
product algebra $A\tens H$ iff $(\chi,\la)$ is a coboundary.
\end{propos}

\proof Again, this is known from \cite{Doi:equ}, but it is instructive to give
a direct and self-contained proof. We begin by showing that any algebra map
$\tilde\gamma:A{}_{\chi'}\lcross H\to A{}_\chi\lcross H$ covariant under the
right coaction of $H$ and restricting to the identity on $A$ is necessarily of
the form
\eqn{tildegamma}{\tilde\gamma(a\tens h)=a\gamma(h\o)\tens h\t,\quad \gamma:H\to
A.}
Just define $\gamma(h)=(\id\tens\eps)\circ\tilde\gamma(1\tens h)$ and check
that
\align{&&\equad a\gamma(h\o)\tens h\t=a\tilde\gamma(1\tens h\o)\uo\eps(
\tilde\gamma(1\tens h\t)\ut)\tens h\t\\
&&=a\tilde\gamma(1\tens h)\uo\eps (\tilde\gamma(1\tens h)\ut\o)\tens
\tilde\gamma(1\tens h)\ut\t
=(a\tens 1)\tilde\gamma(1\tens h)}
where $\tilde\gamma=\nosum \tilde\gamma\uo\tens\tilde\gamma\ut
$ is an explicit notation for the output of $\tilde\gamma$ in $A\tens H$. We
used the assumption of covariance under the right coaction of $H$ for the
second equality. Conversely, it is clear that any linear map $\gamma:H\to A$
defines an $H$-comodule map $\tilde \gamma$. Finally, note under this
correspondence that $\tilde\gamma$ is an isomorphism iff $\gamma$ is
convolution-invertible. Here $\tilde\gamma^{-1}$ is provided in the same way as
(\ref{tildegamma}) with $\gamma$ replaced by the  convolution-inverse
$\gamma^{-1}$.

Next, evaluating the product in $A{}_\chi\lcross H$ and the image of the
product in $A_{\chi'}\lcross H$ we have respectively
\align{ \tilde\gamma(a\tens h)\tilde\gamma(b\tens g)\equad
&&=(a\gamma(h\o)\tens h\t)\cdot(b\gamma(g\o)\tens g\t)\\
&&=a\gamma(h\o) h\t\la(b\gamma(g\o))\chi(h\th\tens g\t)\tens h\fo g\th\\
&&=a (h\o\la^\gamma b)\chi^\gamma(h\t\tens g\o)\gamma(h\th g\t)\tens h\fo
g\th\\
 \tilde\gamma((a\tens h)(b\tens g))\equad&& =\tilde\gamma\left(a (h\o\la'
b)\chi'(h\t\tens g\o)\tens h\th g\t\right)\\
&&=a (h\o\la' b)\chi'(h\t\tens g\o)\gamma(h\th g\t)\tens h\fo g\th}
Assuming these expressions coincide, we apply $\id\tens \gamma^{-1}$ to  both
and multiply in $A$, to conclude that
\[ (h\o\la^\gamma b)\chi^\gamma(h\t\tens g)=(h\o\la' b)\chi'(h\t\tens g).\]
Setting first $g=1$ and then $b=1$, we conclude that
$(\chi',\la')=(\chi^\gamma,\la^\gamma)$ as required. Conversely, it is clear
that if this equality does hold then $\tilde\gamma$ is an algebra homomorphism.
It is clear also that $\tilde\gamma$ preserves the unit $1\tens 1$ iff
$\gamma(1)=1$. \endproof

{}From this cocycle cross product algebra, there are systematic techniques to
generate all the right-handed and dual versions that we might need. For
example, writing the constructions in terms of maps pointing downwards, a
left-right reversal gives the formulae for a right-handed cocycle cross product
algebra, while an up-down reflection gives the formulae for a left-handed
cocycle cross coproduct coalgebra. This is one of the beautiful things one can
do with Hopf algebras, because their axioms are symmetric under these
reflections. Without going into details, it is clear that the necessary data
for the dual construction is a cocycle-coaction
\[ \beta:C\to H\tens C,\quad \psi:C\to H\tens H\]
say where $\psi$ is a cocycle {\em in} $H$ with values {\em from} the coalgebra
$C$. There is the dual notion to (\ref{modalg}), namely that $\beta$ respects
the coproduct of $C$, and there are the axioms of a cocycle-coaction which come
out as
\ceqn{cy-lcomod}{\equad \nosum ((\id\tens\beta)\circ\beta(c\o))(\psi(c\t)\tens
1)=\nosum (\psi(c\o)\tens 1)((\Delta\tens\id)\circ\beta(c\t))  \\
(\eps\tens\id)\circ\beta(c)=c}
\ceqn{2-cy-comod}{ \nquad\nquad \nosum
((\id\tens\psi)\circ\beta(c\o))((\id\tens\Delta)\circ\psi(c\t))=\nosum
(\psi(c\o)\tens 1)(((\Delta\tens\id)\circ\psi(c\t))\\
(\eps\tens\id)\circ\psi(c)=\eps(c)=(\id\tens\eps)\circ\psi(c)}
for all $c\in C$. The resulting cocycle cross coproduct coalgebra
$C^\psi\lcocross H$ is
\eqn{cocyclecoprod}{ \Delta(c\tens h)=\nosum h\o\tens c\t\bo\psi(c\th)\uo
h\o\tens
c\t\bt\tens \psi(c\th)\ut h\t}
and tensor product counit. The cohomology picture likewise goes through: The
group of convolution-invertible linear maps $\gamma:C\to H$ with
$\eps\circ\gamma=\eps$ acts on the pair $(\psi,\beta)$ by
\ceqn{cogauge}{\psi^\gamma(c)=\nosum (\gamma(c\o)\tens
1)((\id\tens\gamma)\circ\beta(c\t))\psi(c\th)(\Delta\gamma(c\fo))\\
\beta^\gamma(c)=(\gamma(c\o)\tens 1)\beta(c\t)(\gamma^{-1}(c\th)\tens 1)}
and the equivalence classes under this are the elements of the nonAbelian
cohomology spaces $\CH^2(C,H)$. All of this is a straightforward dualisation of
the above results for cross product algebras. Now we combine the two
constructions to obtain

\begin{theorem}\cite{Ma:mor}\cite{MaSoi:bic} Let $H$ and $A$ be bialgebras,
$(\chi,\ra)$ a right handed cocycle-action of $H$ on $A$ and $(\psi,\beta)$ a
cocycle-coaction of $A$ on $H$. If in addition we have
\ceqn{cycobic}{ \nosum \psi(h\o)\Delta (a\ra h\t)= \nosum \left((a\o\ra h\o)
h\t\bo\tens a\t\ra h\t\bt\right)\psi(h\th)\\
\eps(a\ra h) =\eps(a)\eps(h),\qquad \beta(1)=1\tens 1\\
\nosum\beta(h\o g\o)(\chi(h\t\tens  g\t)\tens1)=\nosum \chi(h\o\tens
g\o)(h\t\bo\ra g\t) g\th\bo\tens h\t\bt g\th\bt\\
\nosum h\o\bo (a\ra h\t)\tens h\o\bt= \nosum (a\ra h\o) h\t\bo\tens h\t\bt}
and $(\chi,\psi)$  obey a compatibility condition
\ceqn{cycobicd}{\nosum \psi(h\o g\o)\Delta\chi(h\t\tens  g\t)=\nosum
(\chi(h\o\tens  g\o)(h\t\bo\ra g\t)
g\th\bo(\psi(h\th)\uo\ra g_{(4)}) g_{(5)}\bo\\
\qqquad\qquad \tens \chi(h\t\bt\tens
g\th\bt)(\psi(h\th)\ut\ra g_{(5)}\bt))\psi(g\s6)\\
\eps(\chi(h\tens g))=\eps(h)\eps(g),\quad \psi(1)=1\tens 1}
then the cocycle cross product algebra and cross coproduct coalgebra fit
together to form the {\em cocycle bicrossproduct}  bialgebra
$H{}^{\psi}\cobicross_{\chi}A$.
\end{theorem}

These Hopf algebra constructions may appear unfamiliar, but this is exactly the
machinery which recovers the familiar theory of central extensions and unifies
them with other cocycle constructions. The central extensions as in
(\ref{ext-gp})--(\ref{ext-gp-E}) are given in terms of the respective group
algebras by $\psi,\beta,\ra$ all trivial and only $\chi:G\times G\to M$
non-trivial, where $M$ is an Abelian group and $A=\C M$.

On the other hand, we answer also the question of when the cocycle cross
products $\C G\rcross_\chi\C(M)$ corresponding to anomalous quantisations of
homogeneous spaces, are Hopf algebras. Thus, let $G$ be a finite group and $A$
an algebra. The condition that we have a (right-handed) cocycle cross product
algebra $\C G\rcross_\chi A$ is that we have maps $\chi:G\times G\to A$,
$\ra:A\times G\to A$ obeying
\ceqn{rgpcy}{1\ra u=1,\quad (ab)\ra u=(a\ra u)(b\ra u)\\
\chi(u,v)((a\ra u)\ra v)=(a\ra (uv))\chi(u,v),\quad a\ra 1=a\\
\chi(uv,w)(\chi(u,v)\ra w)=\chi(u,vw)\chi(v,w),\quad \chi(e,u)=\chi(u,e)=1}
which, in the case $A$ commutative would just be the usual notion of a group
two-cocycle in $Z^2(G,A)$. The resulting cross product algebra is
\[ (u\tens a)(v\tens b)=uv\tens \chi(u,v)(a\ra v)b,\quad 1=e\tens 1\]
It we set $A=\C(M)$ for example, we have exactly a right-handed version of the
anomalous cross product quantisation (\ref{cocyquant}) discussed at the start
of the section.

So for this to become a Hopf algebra along the lines above, we need $A$ to be a
Hopf algebra, and a cocycle with values {\em from} $G$ in the form of a map
$\psi:G\to A\tens A$. This means nothing other than
\ceqn{g2-cocycle}{\psi(u)_{23}(\id\tens\Delta)\psi(u)
=\psi(u)_{12}(\id\tens\Delta)\psi(u)\nonumber\\
(\eps\tens\id)\psi(u)=(\id\tens\eps)\psi(u)=1}
for all $u\in G$. To keep things simple, we take coaction $\beta$ to be
trivial, so that we can concentrate on the cocycle $\psi$. The cocycle cross
coproduct (\ref{cocyclecoprod}) then reduces to
\eqn{gpcocycoprod}{\Delta(u\tens a)=\nosum u\tens \psi(u)\uo a\o\tens u\tens
\psi(u)\ut a\t,\quad \eps(u\tens a)=\eps(a).}
Finally, we ask that these give a Hopf algebra, $\C G^\psi\cobicross_\chi A$.
According to the general theory above, the conditions for this
are\cite{Ma:mor}\cite{MaSoi:bic}
\eqn{cyGAa}{ \eps(a\ra u)=\eps(a),\quad \psi(u)\Delta (a\ra u)=((\Delta a)\ra
(u\times u))\psi(u)}
\ceqn{cyGAb}{\psi(uv)\Delta\chi(u,v)=(\chi(u,v)\tens \chi(u,v))(\psi(u)\ra
(v\times v))\psi(v)\nonumber\\
\quad \eps (\chi(u,v))=1,\qquad \psi(e)=1\tens 1.}

As well as describing interesting possibilities for quantum physics, this
construction has important algebraic examples too. For the very simplest case
we let $G=\Z_2$ the group with two elements $\{e,w\}$ say and $w^2=1$. Then our
family (\ref{g2-cocycle}) becomes just one 2-cocycle $\psi(w)\in A\tens A$, our
action $\ra$ just becomes one algebra automorphism $T=\ra w:A\to A$ and $\chi$
just becomes one element $x=\chi(w,w)\in A$. The cocycle conditions above
become (\ref{g2-cocycle}) and
\[  \psi(\Delta T(a))=((T\tens T)\Delta a)\psi,\quad \eps\circ T(a)=\eps(a)\]
\[ x^{-1}ax=T^2(a),\quad T(x)=x,\quad \Delta x=(x\tens x)((T\tens
T)(\psi))\psi,\ \eps(x)=1.\]
So this data defines a cocycle bicrossproduct Hopf algebra
$\C\Z_2{}^\psi\cobicross_\chi A$ generated by $A$ as a subHopf algebra and one
additional generator $w\equiv w\tens 1$ with
\[w^{-1}aw=T(a),\ w^2=x,
\quad\Delta w=(w\tens w)\psi,\ \eps w=1,\ Sw=w U Sx\]
where $U=\psi\uo S\psi\ut$.

\begin{example} Let $A=U_q(sl_2)$ in the standard conventions and let
$\psi=\CR$ its quasitriangular structure, $x=\nu^{-1}$ the inverse of its
ribbon element in the sense of\cite{ResTur:rib}, and $T$ the algebra
automorphism
\[ T(H)=-H,\quad T(X_\pm)=-q^{\pm\h}X_\mp.\]
This data obeys the conditions above. The resulting cocycle bicrossproduct Hopf
algebra is generated by $U_q(sl_2)$ and one element $w$ adjoined with relations
\[ w^{-1}aw=T(a),\ w^2=\nu^{-1},\quad \Delta w=(w\tens w)\CR,\ \eps w=1,\ Sw=w
q^{-H}.\]
\end{example}
This extended Hopf algebra is a variant of the {\em quantum Weyl group} of
$U_q(sl_2)$ and $T$ is a variant of Lusztig's automorphism\cite{Lus:roo}. This
example and the general situation for $U_q(g)$ are in \cite{MaSoi:bic}. We have
given here a new and more explicit version.

\section{Remark on trivial quantum principal bundles}

A fascinating aspect of quantum groups is that they can be interpreted either
as like enveloping algebras, or (the same Hopf algebras), as like algebras of
functions. This means that the cross product quantisations of Sections~2--4 in
which the group or quantum group generates the momentum symmetry, can appear
instead as like the algebra of functions on some geometrical object. We
demonstrateed exactly this for the $\hbar\to 0$ limit of our Example~2.3, where
$\C[x]\bicross \C[p]\isom \C(\R\rcross\R)$. Here $\R\rcross\R$ is the phase
space and is curved. Now as soon as $\hbar\ne 0$ our quantum algebra of
observables $\C[x]\bicross \C[p]$ is non-commutative and so it cannot literally
be the algebra of functions on any ordinary space. Rather, it is by definition
the algebra of functions on some {\em quantum space}. In this section, we want
to elucidate exactly what is this quantum geometrical structure. In fact, the
general machinery which we need is that of quantum group gauge theory as
introduced in \cite{BrzMa:gau} and this section is intended as an introduction
to this work.

We will see from this point of view that the cross product algebras in
Section~2, and in Section~4 with $\chi$ convolution-invertible are trivial
quantum principal bundles. They are geometrically trivial according to any
usual notion of topology from the base manifold etc, but we still know from
Proposition~4.2 that they are classified by the nonAbelian cohomology
$\CH^2(H,A)$, which appears now as a new kind of topological invariant present
only when the principal bundle is a quantum space. These are the same cocycles
$\chi$ which, from the quantisation point of view, relate to anomalies.

The explanation of our point of view is very simple. Recall that a usual
trivial principal bundle $P$ over a manifold $M$ and with structure group $G$
is a presentation of the base manifold as a homogeneous space $M=P/G$ and a
global group co-ordinate chart $P\to G$. This is surjective and intertwines the
action on $P$ with the right action of $G$ on itself by multiplication. Along
with the canonical projection, it provides the global trivialisation $P\isom
M\times G$. Now it is easy to reformulate these notions in an algebraic way.
Firstly, we require a right coaction $\C(P)\to \C(P)\tens\C(G)$ such that
$\C(M)$ appears as the fixed point subalgebra $\C(P)^{\C(G)}$. In addition, we
need a global trivialisation map  $j:\C(G)\to\C(P)$ which should be injective
and an intertwiner for the right coaction on $\C(P)$ and the right coaction of
$\C(G)$ on itself by comultiplication. For a non-trivial principal bundle one
does not have this global trivialisation $j$ but other weaker conditions, such
as freeness of the action and local exactness conditions\cite{BrzMa:gau}.

Thus the abstract model of a trivial principal bundle is to replace $\C(G)$ by
$H$, a Hopf algebra; $\C(P)$ by $E$, an algebra on which $H$ coacts by an
algebra homomorphism $\beta:E\to E\tens H$, and $\C(M)$ by an algebra $A$
identified with the fixed point subalgebra
\[ E^{H}=\{e\in E|\ \beta(e)=e\tens 1\}\subseteq E.\]
This data is called an {\em extension} of an algebra $A$ by a Hopf algebra $H$.
Next, the idea of a global trivialisation means a linear map $j:H\to E$ which
should be covariant under the right coaction of $H$. We do not insist that it
is an algebra
homomorphism but it should be convolution-invertible and obey $j(1)=1$. In this
case, the extension is said to be {\em cleft}. On the other hand, this data
$(A,E,H,j)$ is exactly what we have used in \cite{BrzMa:gau} as the starting
point of quantum group gauge theory. This notion of extensions is also an
important idea in Galois theory, where one is interested in extensions of
fields. So this area of number-theory is also unified with the present
considerations of quantum mechanics and gauge theory. Some other
geometrically-motivated results at the Hopf algebraic level are in
\cite{Sch:rep}.

On the other hand, it is not hard to see that the cocycle cross products
$E=A{}_\chi\lcross H$ of Section~4 with convolution-invertible cocycle $\chi$
are always cleft extensions. Conversely, every cleft extension of an algebra
$A$ by a Hopf algebra $H$ is isomorphic to a convolution-invertible cocycle
cross product. One defines $\chi:H\tens H\to E$ and $\la:H\tens A\to E$ by
\ceqn{cleftchi}{j(h\o)j(g\o)j^{-1}(h\t g\t)=\chi(h\tens g)\nonumber\\
j(h\o)i(a)j^{-1}(h\t)=h\la a}
where $i:A\hookrightarrow E$ explicitly denotes the inclusion of $A$ in $E$ as
fixed point subalgebra. It is clear that $(\chi,\la)$ form a trivial cocycle in
$\CH^2(H,E)$ as the coboundary of $j$. On the other hand,  $H$-covariance of
$j$ gives at once that the images of these maps $\chi,\la$ are in the
fixed-point subalgebra, i.e. we they are actually maps $H\tens H\to A$ and
$H\tens A\to A$ as required. They still obey the cocycle conditions but now
viewed as an element of $\CH^2(H,A)$. As such they are not necessarily a
coboundary since $j$ itself need not map to $A$. We then build the cocycle
cross product $A{}_\chi\lcross H$ on the vector space $A\tens H$ and verify
that $A\tens H\to E$ given by $a\tens h\mapsto i(a)j(h)$ is an isomorphism. In
this way, the inequivalent cleft extensions are in one-one correspondence with
the nonAbelian convolution-invertible cohomology classes in $\CH^2(H,A)$. This
is all fairly standard by now; one can see \cite{Doi:equ} and also
\cite{BlaMon:cro}. Note that the notion of equivalence of extensions used here
is the same as in Proposition~4.2, namely an isomorphism which is covariant
under the coaction of $H$ and restricts to the identity on $A$. In the gauge
theory picture these are exactly gauge transformations $\gamma:H\to A$ in the
sense described in \cite{BrzMa:gau}.

Note that in the classical setting,  we assume that $E$ is commutative and that
$j$ is an algebra map. The first of these forces any possible action $\la $ to
be trivial and the second forces any possible cocycle $\chi$ to be trivial
also. Hence $E\isom A\tens H$ as an algebra in this classical situation. This
is why our nonAbelian cohomology does not enter the scene in the usual
classical theory of trivial principle bundles. On the other hand, as soon as we
go beyond this conventional setting, we have the possibility of non-trivial
cocycles and correspondingly new quantum numbers provided by the cohomology
classes in $\CH^2(H,A)$.

\section{NonAbelian cohomology and twisting}

In this section we make a connection between the nonAbelian cohomology spaces
$\CH^2(H,A)$ above and Drinfeld's theory of twisting\cite{Dri:qua}. From a
mathematical standpoint, this is the modest new result of the paper.

Drinfeld introduced his process of twisting for quasi-Hopf algebras. These are
only coassociative up to conjugation by a 3-cocycle $\phi$, obeying
\ceqn{3-cocycle}{{}\nquad(1\tens \phi)((\id\tens\Delta\tens \id)\phi)(\phi\tens
1)=((\id\tens \id\tens \Delta)\phi)((\Delta\tens \id\tens \id)\phi)\nonumber\\
(\id\tens\eps\tens\id)\phi=1\tens 1.}
In this case, twisting consists of conjugating the coproduct by an arbitrary
invertible element $\psi$, say, in $H\tens H$. However, it is easy to
see\cite{GurMa:bra}, that if $\psi$ obeys a further condition
\ceqn{2-cocycle}{ (1\tens \psi)(\id\tens\Delta) \psi=(\psi\tens
1)(\Delta\tens\id)\psi,\quad (\eps\tens\id)\psi=1}
then
 \[ \Delta_\psi =\psi(\Delta\ )\psi^{-1},\quad \CR_\psi=\psi_{21}\CR
\psi^{-1},\quad S_\psi=U(S\ )U^{-1}\]
remains a quasitriangular Hopf algebra, which we denote $H_\psi$. Here
$U=\nosum \psi\uo (S \psi\ut  )$. Of course, the result works also if $H$ is
only a bialgebra. Recently cf\cite{Ogi:zdi}, O. Ogievetsky has pointed out that
our bicrossproduct Hopf algebras in Example~2.3 provide a simple example of
twisting:

\begin{example} The bicrossproduct Hopf algebra $H=\C[x]\bicross \C[p]$ is
triangular and is twisting-equivalent to $U(b_+)$ where $b_+$ is the
two-dimensional solvable Lie algebra. The triangular structure and twisting
cocycle are
\[ \psi=e^{p\tens x\over \hbar},\quad \CR=e^{x\tens p\over\hbar}e^{-{p\tens
x\over\hbar}}.\]
\end{example}
\proof We first change variables to $ X=e^{\sB x}-1$ so that
$p\la  X=\hbar (1-e^{-\sB x}){\del\over\del x}  X=\hbar \sB(1-e^{-\sB x})e^{\sB
x}=\hbar \sB  X$. In terms of this generator $ X$ the bicrossproduct structure
in Example~2.3 becomes
\cmath{ [p,   X]=\hbar \sB    X\\
\Delta    X= X\tens 1+1\tens  X+X\tens X,\quad \Delta  p=1\tens   p+  p\tens
(1+ X)^{-1}.}
The algebra here is just the enveloping algebra of the Lie algebra $b_+$ with
generators $p,  X$. Its coproduct is not that of an enveloping algebra. But
since $[  p\tens   x,p\tens 1]=0=[p\tens x,1\tens X]$ while $[p\tens x,1\tens
p]=-\hbar p\tens X(1+X)^{-1}$ and $[p\tens x,X\tens 1]=\hbar \sB X\tens x$ we
see at once that
\[\psi( X\tens 1+1\tens  X)\psi^{-1}=\Delta  X,\quad  \psi(p\tens 1+1\tens
p)\psi^{-1}=\Delta p.\]
This means that the coalgebra is the twisting of $U(b_+)$. One can easily check
the assertion for the counit also. Moreover, since the latter Hopf algebra has
the trivial quasitriangular structure $1\tens 1$, we deduce that
$\CR=\psi_{21}\psi^{-1}$ is a quasitriangular structure for the bicrossproduct
Hopf algebra. It is triangular in the sense $\CR_{21}\CR=1$. \endproof

As far as I know, \cite{GurMa:bra} is perhaps the first place where the
condition (\ref{2-cocycle}) was explicitly formulated (those in
\cite{Res:mul}, for example, being stronger) while remaining of course
no more than a special case of Drinfeld's ideas in \cite{Dri:qua}.  We
observe now that this condition is exactly a special case of the
cocycles $\psi$ in Section~4, namely $\CH^2(\C,H)$ in the notation
there.  Let us add to these the conditions of a 1-cocycle. Just as in
the dual version of Proposition~4.1, this is now an element $\gamma\in
H$ such that \eqn{1-cocycle}{ \gamma\tens\gamma=\Delta\gamma,\quad
\eps\gamma=1} i.e., a group-like element. Then we have as a special
case of (\ref{cogauge}):

\begin{propos} Let $H$ be a bialgebra or Hopf algebra. If $\gamma\in H$ is an
invertible element with $\eps\gamma=1$ then
$\del\gamma=(\gamma\tens\gamma)\Delta\gamma^{-1}$ is a 2-cocycle in $H$. We say
that it is a {\em coboundary}. More generally, if $\psi$ is a 2-cocycle then
\[ \psi^\gamma=(\gamma\tens\gamma)\psi\Delta\gamma^{-1}\]
is also a 2-cocycle. We say that it is cohomologous to $\psi$. The non-Abelian
cohomology space $\CH^2(\C,H)$ is the 2-cocycles in $H$ modulo such
transformations.
\end{propos}
\proof For completeness, we give here a direct proof of the transformation of
cocycles. Thus
\align{&&\equad (1\tens \psi^\gamma)(\id\tens\Delta)\psi^\gamma\\
&&=(1\tens(\gamma\tens\gamma)\psi\Delta\gamma^{-1})(\id\tens\Delta)
((\gamma\tens\gamma)\psi\Delta\gamma^{-1})\\
&&=(\gamma\tens\gamma\tens\gamma)(1\tens
\psi\Delta\gamma^{-1})(\id\tens\Delta)(1\tens\gamma \psi\Delta\gamma^{-1})\\
&&=(\gamma\tens\gamma\tens\gamma)((1\tens
\psi)(\id\tens\Delta)\psi)(\id\tens\Delta)\Delta\gamma^{-1}\\
&&=(\gamma\tens\gamma\tens\gamma)((\psi\tens 1)(\Delta\tens
\id)\psi)(\id\tens\Delta)\Delta\gamma^{-1}\\
&&=(\gamma\tens\gamma\tens 1)(\psi\Delta\gamma^{-1}\tens
1)((\Delta\tens\id)(\gamma\tens\gamma)\psi)(\Delta\tens\id)\Delta\gamma^{-1}\\
&&=(\psi^\gamma\tens 1)(\Delta\tens \id)\psi^\gamma}
as required. The other facts are clear.
\endproof

For example, \cite[Lemma~2.2]{GurMa:bra} says in these terms that
\[ \psi\sim (S\tens S)(\psi_{21}^{-1}).\]
Likewise, $\CR$ itself regarded as a 2-cocycle is always cohomologous to
$\CR_{21}^{-1}$. The twisting in these cases is clearly just the opposite Hopf
algebra.

\begin{propos} Let $\psi,\psi'$ be 2-cocycles. The Hopf algebras given by
twisting by them are isomorphic via an inner automorphism if $\psi,\psi'$ are
cohomologous. I.e. there is a map from $\CH^2(\C ,H)$ to the set of twistings
of $H$ up to inner automorphism. In particular, if $\psi$ is a coboundary then
twisting by it can be undone by an inner automorphism.
\end{propos}
\proof We suppose that $\psi,\psi'$ are cohomologous in the sense of
Proposition~6.2. This means $\psi'=(\gamma\tens\gamma)\psi\Delta\gamma^{-1}$
for some invertible $\gamma\in H$. Then we have $\Delta_\psi'(h)=\psi'(\Delta
h)\psi'^{-1}=(\gamma\tens\gamma)\psi(\Delta\gamma^{-1})(\Delta h)
(\Delta\gamma)\psi^{-1}(\gamma^{-1}\tens\gamma^{-1})
=(\gamma\tens\gamma)(\Delta_\psi(\gamma^{-1} h\gamma))(\gamma^{-1}\tens
\gamma^{-1})$. As $\gamma(\ )\gamma^{-1}$ is an inner automorphism of the
algebra structure, we see that it defines now a bialgebra isomorphism
$H_\psi'\to H_\psi$. Hence it is also a Hopf algebra isomorphism in the case
where $H$ has an antipode. One can also see this directly from the formulae
given for the antipode after twisting. Finally, if $H$ is quasitriangular
then $\CR_\psi'=\psi'_{21}\CR \psi'^{-1}=(\gamma\tens\gamma)\psi_{21}
(\Deltaop\gamma^{-1})\CR(\Delta\gamma)\psi^{-1}(\gamma^{-1}\tens
\gamma^{-1})=(\gamma\tens\gamma)\CR_\psi(\gamma^{-1}\tens\gamma^{-1})$ using
the axioms of a quasitriangular structure. So the induced isomorphism maps
the quasitriangular structures too if these are present. \endproof

This says that the process of twisting can only give a genuinely new Hopf
algebra if the cocycle $\psi$ used to twist is cohomologically non-trivial. For
example, if $\CH^2$ is trivial for a Hopf algebra then all twists are
isomorphic. On the other hand, this gloomy possibility does not seem to occur
very often and
the twisting process does generally enable one to obtain new quasitriangular
Hopf algebras from old.

We remark that all of these 1-cocycles, 2-cocycles and also Drinfeld's
3-cocycles are nicely described by the following general formulae for
n-cocycles $Z^n(\C,H)$. On the other hand, though a reasonable coboundary map
$\del$ is defined, it obeys $\del^2=1$ only in a subtle sense in which the Hopf
algebra etc is modified after the first application of $\del$. This is already
evident in Drinfeld's theory of quasi-Hopf algebras and only in the case of
1-cocycles and 2-cocycles does this `feedback' not arise. With this caveat, we
can still proceed with some reasonable formulae. Thus, let $H$ be a bialgebra
or Hopf algebra. We let
\eqn{face}{\Delta_i:H^{\tens n}\to H^{\tens n+1},\quad
\Delta_i=\id\tens\cdots\tens\Delta\tens\cdots\tens\id}
where $\Delta$ is in the $i$'th position. Here $i=1,\cdots,n$ and we add to
this the conventions $\Delta_0=1\tens(\ ) $ and $\Delta_{n+1}=(\ )\tens 1$ so
that $\Delta_i$ are defined for $i=0,\cdots, n+1$. We define an $n$-cochain
$\psi$ to be an invertible element of $H^{\tens n}$ and we define its {\em
coboundary} as the  $n+1$-cochain
\eqn{cobound}{ \del \psi=\left(\prod_{i=0}^{i\ {\rm even}} \Delta_i
\psi\right)\left(\prod_{i=1}^{i\ {\rm odd}} \Delta_i \psi^{-1}\right)}
where the even $i$ run $0,2,\cdots,$ and the odd $i$ run $1,3,\cdots,$ and the
products are each taken in increasing order. We also write $\del \psi\equiv
(\del_+\psi)(\del_-\psi^{-1})$ for the separate even and odd parts. An
$n$-cocycle {\em in} a Hopf algebra or bialgebra is an invertible element
$\psi$ in $H^{\tens n}$ such that $\del \psi=1$. Finally, we assume that our
cochains or cocycles are counital in the sense $\eps_i \psi=1$ for all
$\eps_i=\id\tens\cdots\tens\eps\tens\cdots\tens\id$. One can also posit the
formula
\eqn{psitrans}{ \psi^\gamma=(\del_+\gamma)\psi(\del_-\gamma^{-1})}
for transformation of a cocycle by a cochain of one degree lower, with the
caveat about feedback mentioned above when $n\ge 3$.

These formulae recover of course the 1-, 2-, and 3-cocycles above and their
transformation properties. Moreover, they reduce for $H=\C(G)$ to the usual
theory of group cocycles. We recall that a usual group $n$-cochain is a
point-wise invertible function $\psi:G\times G\cdots\times G\to \C $ and has
coboundary
\[ (\del \psi)(u_1,u_2,\cdots
,u_{n+1})=\prod_{i=0}^{n+1}\psi(u_1,\cdots,u_iu_{i+1},\cdots
,u_{n+1})^{(-1)^i}\]
where, by convention, the first $i=0$ factor is $\psi(u_2,\cdots, u_{n+1})$ and
the last $i=n+1$ factor is $\psi(u_1,\cdots ,u_n)^{\pm 1}$. One has $\del^2=1$
and the group cohomology $\CH^n(G,\C )$ is then defined as the group of
$n$-cocycles modulo multiplication by coboundaries of the form $\del(\ )$. On
the other hand, these `classical' formulae are equally interesting when  the
commutative Hopf algebra $\C(G)$ is replaced by something non-commutative, even
the group algebra $\C G$ or an enveloping algebra $U(g)$.

For completeness, we now give the dual version of all these constructions, for
cocycles $Z^n(H,\C)$. Thus an $n$-cochain {\em on} a bialgebra or Hopf algebra
$H$ is a linear functional $\chi:H^{\tens n}\to \C $ which is invertible in the
convolution algebra and unital in the sense $\chi(h_1\tens \cdots\tens
1\tens\cdots\tens h_{n-1})=\eps(h_1)\cdots\eps(h_{n-1})$, for 1 in any
position. The
coboundary $\del\chi$ is an $n+1$-cochain of the same form as (\ref{cobound})
but with the product in the convolution algebra,
\[ \del\chi=\left(\prod_{i=0}^{i\ {\rm
even}}\chi\circ\cdot_i\right)\left(\prod_{i=1}^{i\ {\rm
odd}}\chi^{-1}\circ\cdot_i\right)\]
where $\cdot_i$ is the map that multiplies $H$ in the $i,i+1$ positions. The
convention is $\chi\circ\cdot_0=\eps\tens\chi$ and
$\chi\circ\cdot_{n+1}=\chi\tens\eps$. For example, a 1-coboundary is
\[ \del\chi(h\tens g)=\nosum \chi(g\o)\chi(h\o)\chi^{-1}(h\t g\t)\]
so that a 1-cocycle just means an algebra homomorphism $H\to \C$. 3-cocycles
provide of course dual quasiHopf algebras as studied in \cite{Ma:tan}.
Likewise, 2-cocycles provide the theory of twisting in the dual formulation.
Thus if $\chi$ is a  2-cocycle on a Hopf algebra $H$ in the sense $\chi(h\tens
1)=\eps(h)=\chi(1\tens h)$ and
\[\nosum\chi(g\o\tens f\o)\chi(h\tens g\t f\t)=\nosum\chi(h\o\tens g\o)\chi(h\t
g\t\tens f)\]
then we obtain another Hopf algebra $H_\chi$ by twisting the product and
antipode according to
\[ h\cdot_\chi g=\nosum \chi(h\o\tens g\o)h\t g\t \chi^{-1}(h\th\tens g\th)\]
\[ S_\chi(h)=\nosum U(h\o)Sh\t U^{-1}(h\th),\quad U(h)=\nosum \chi(h\o\tens
Sh\t).\]
If $H$ is dual-quasitriangular then
\[ {\CR}_\chi(h\tens g)=\nosum \chi(g\o\tens h\o)\CR(h\t\tens g\t)
\chi^{-1}(h\th\tens g\th)\]
is a dual-quasitriangular structure on the twisted Hopf algebra.

We have seen in Section~4 that 2-cocycles provide on the one hand central
extensions, as encountered in anomalous quantum theories, and on the other hand
they provide also dual-twistings. It seems likely that this connection could be
exploited in the study of anomalies. Moreover, we have the possibility too of
even more general anomalies than those usually encountered and which we might
expect to show up in $q$-deformed physics. On the mathematical side, one would
like to see these nonAbelian cohomology spaces connected more closely with
cohomological ideas in the abstract deformation theory of Hopf algebras and
their generalisations, as started in
\cite{GerSch:bia}\cite{ShnSte:cob}\cite{MarSta:def} and elsewhere. A close
connection could be expected since deformation theory and cohomology were part
of the original background behind Drinfeld's twisting construction.

\baselineskip 12pt
\itemsep 0pt


\end{document}